\documentclass[9pt,twocolumn,twoside]{opticajnl}
\journal{opticajournal} 
\doi{} 
\setboolean{shortarticle}{false}

\usepackage{silence}
\usepackage{amsmath, bm, mathcomp}
\usepackage{changebar}
\usepackage{lineno}

\title{Theory of interferometric background suppression for highly sensitive electro-optic sampling}

\author[1]{Hiroaki Ito}
\author[1]{Daiki Okazaki}
\author[1]{Shigeki Tokita}
\author[2,3,*]{Masato Ota}

\affil[1]{Institute for Chemical Research, Kyoto University, Gokasho, Uji, Kyoto 611-0011, Japan}
\affil[2]{National Institute for Fusion Science, 322-6 Oroshi-cho, Toki, Gifu 509-5292, Japan}
\affil[3]{SOKENDAI (The Graduate University for Advanced Studies), 322-6 Oroshi-cho, Toki, Gifu 509-5292, Japan}

\affil[*]{ota.masato@nifs.ac.jp}

\begin{abstract}
In electro-optic (EO) sampling, increasing the probe power enhances the EO signal, but the usable power can be limited by photodetector saturation caused by the much larger probe background independent of the applied electric field.
Here, we propose a method for suppressing this background through destructive interference between an EO-modulated probe and an unmodulated reference beam.
Using a Jones-matrix model, we compare five detection configurations and identify balanced detection as suitable for maintaining a linear weak-field response under strong background suppression.
This suppression allows higher input optical power to be used without saturating the photodiodes, thereby enhancing the EO signal.
Under a per-photodiode saturation constraint, the optimized signal enhancement scales approximately as the square root of the total input power.
For a total input power of $5~\mathrm{W}$ and a saturation limit of $5~\mathrm{mW}$ per photodiode, the optimized configuration yields an approximately 12-fold signal enhancement over conventional balanced detection.
\end{abstract}

\setboolean{displaycopyright}{false} 

\begin{document}

\maketitle

\section{Introduction}
Electro-optic (EO) sampling detects electric fields through the field-induced birefringence of an EO crystal, which produces a phase retardation in an optical probe.
EO sampling was initially developed for subpicosecond measurements of electrical transients \cite{Valdmanis1983} and subsequently extended to coherent detection of far-infrared fields \cite{Auston1984} and to free-space sampling of terahertz (THz) pulses \cite{Wu1995}. 
It is now used for field-resolved measurements from THz to optical frequencies \cite{BeneaChelmus2025}. 
By providing access to both the amplitude and phase of the electric field, EO sampling enables the determination of complex optical constants in time-domain spectroscopy \cite{Duvillaret1996,Koch2023}. 
Its applications also include field-resolved molecular spectroscopy \cite{Pupeza2020, Okazaki2026} and studies of ultrafast current dynamics through THz emission spectroscopy \cite{Kampfrath2013,Sotome2021}. 
Beyond spectroscopy, EO sampling has been used to characterize transient electric fields generated in laser--plasma interactions \cite{Pompili2018} and electric fields associated with ultrashort relativistic electron bunches \cite{Berden2007,Ota2021,Ota2022,Huang2024}. 
It has further been extended to direct sampling of vacuum electric-field fluctuations \cite{Riek2015}.

Detecting weak electromagnetic fields remains challenging, particularly when the available field strength is limited by source output, sample absorption, or unwanted sample excitation. 
The EO-induced modulation can be increased by using a crystal with a larger EO coefficient or a longer interaction length.
However, the available EO materials are limited, and increasing the interaction length involves a trade-off between sensitivity and bandwidth because of absorption and dispersive effects, including velocity mismatch between the THz field and the optical probe \cite{Gallot1999,BeneaChelmus2025}.
Increasing the probe power raises both the EO-induced signal and the much larger field-independent background, which can saturate the photodetector before the useful signal becomes sufficiently large. 
Preferential suppression of this background before photodetection therefore allows higher probe powers and hence larger EO signals while maintaining linear detector operation~\cite{beckh2021}.

Several optical readout strategies have been developed for this purpose. 
Spectral post-filtering can reduce shot noise by rejecting probe spectral components that contribute little to the EO signal \cite{Porer2014}. 
Polarization-based approaches include operation near the extinction point of crossed polarizers \cite{Jiang1999,Xu2008}, subtraction of signals acquired at opposite optical biases \cite{Brunner2014,Johnson2014}, balanced phase-offset detection \cite{Pan2014}, and selective attenuation of the dominant field-independent polarization component \cite{Kato2018Polarization}. 
In the ideal weak-field limit, attenuating only the field-independent amplitude by a factor \(b\) (\(0<b<1\)) reduces the background intensity by a factor of \(b^2\) and the linear EO signal by a factor of \(b\), thereby enhancing the background-normalized modulation by a factor of \(1/b\) at fixed incident probe power \cite{Kato2018Polarization}. 
More recently, Stromberg and Heilweil experimentally demonstrated amplification of EO-detected THz signals using optical-bias-based detection schemes \cite{Stromberg2024}.
In practice, imperfect polarization extinction, residual crystal birefringence, and scattered light limit the achievable suppression and determine the optimal operating point \cite{Jiang1999,Xu2008,Brunner2014}. 
These limitations motivate complementary methods for suppressing the field-independent probe component.

An optical interferometer offers such a route by canceling this background through destructive interference.
A prominent example is laser-interferometric gravitational-wave detection, where destructive interference suppresses the strong optical carrier at the dark output port while differential phase shifts produce a measurable signal \cite{Abbott2016}. 
Related approaches have been applied to background suppression in absorption and vibrational spectroscopy \cite{Tomberg2019,Song2022}. 
Interferometric EO readout has also been demonstrated using a modified Sagnac interferometer to detect THz-induced phase shifts \cite{Clough2009}, cross-polarized spectral-domain interferometry to extend the dynamic range \cite{Ibrahim2016}, and integrated Mach--Zehnder THz sensors \cite{Wilke2024}. 
These implementations primarily employ interferometry for phase readout or dynamic-range extension; in contrast, interferometric suppression of the field-independent probe background before polarization analysis in bulk-crystal EO sampling has received comparatively limited attention.
The sensitivity-enhancement mechanism shares the same general principle as polarization filtering~\cite{Kato2018Polarization}, namely, suppression of the field-independent background before photodetection.
Unlike purely polarization-based readout, however, interference with an independent, unmodulated reference beam also enables EO-induced optical phase shifts to be converted into intensity changes even when the probe polarization state remains unchanged.

In this work, we theoretically investigate an interferometric background-suppression scheme for EO sampling under photodetector-saturation constraints.
Using a Jones-matrix formalism, we compare five representative EO detection configurations and examine their weak-field responses as functions of amplitude and phase imbalances.
The analysis shows that balanced detection retains a first-order response to the THz electric field at finite amplitude imbalance, while its differential output eliminates even-order contributions; complete background cancellation, in contrast, leaves a cubic leading response.
We further determine the amplitude imbalance that maximizes the differential EO signal for different phase imbalances under photodetector-saturation constraints and quantify the resulting enhancement relative to conventional balanced detection.

\section{Theoretical model}
\label{sec:theory}

\begin{figure}[!t]
    \centering
    \includegraphics[width=\linewidth]{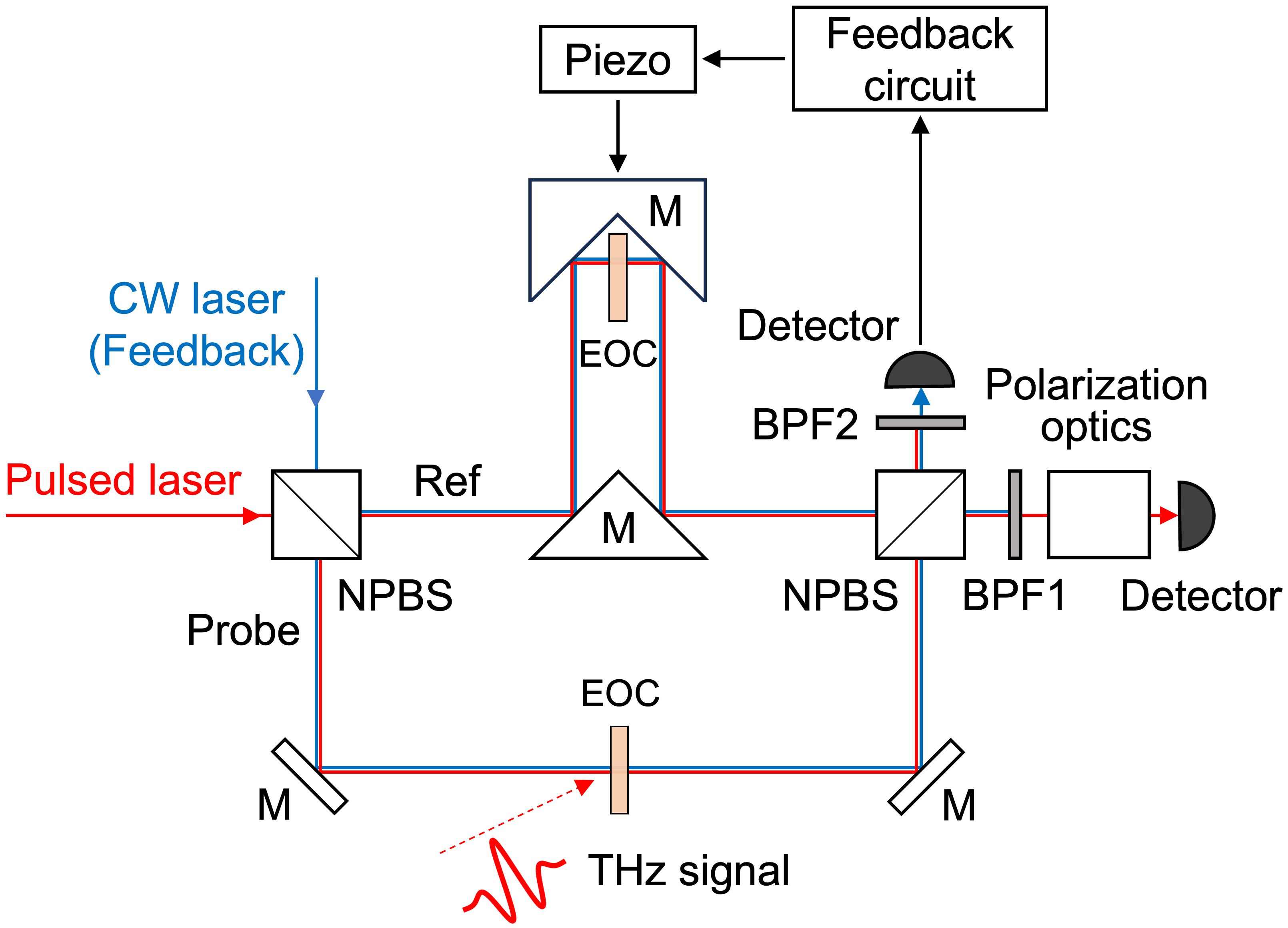}
    \caption{
    Schematic of the proposed interferometric background-suppression scheme for electro-optic (EO) sampling based on a Mach--Zehnder interferometer.
    M, mirror; NPBS, non-polarizing beam splitter; EOC, electro-optic crystal; BPF, band-pass filter; CW laser, continuous-wave laser.
    The CW laser is introduced to stabilize the destructive-interference condition.
    The optical power at the dark port is monitored by the detector behind BPF2 and used as a feedback signal for the feedback circuit, which controls the piezoelectric stage (Piezo) to maintain destructive interference.
    BPF1 blocks the CW laser, whereas BPF2 blocks the pulsed laser.
    The detection configurations after the recombination NPBS are shown in Figs.~\ref{Figure2} and \ref{Figure4}.
    }
    \label{Figure1}
\end{figure}

We analyze the proposed background-suppressed EO sampling scheme using the Jones-matrix formalism, assuming the Mach--Zehnder interferometer shown in Fig.~\ref{Figure1} for destructive interference between the probe and reference beams.
The model assumes that the probe acquires the EO-induced retardation while the reference remains unmodulated by the measured field, and that the probe and reference beams are perfectly matched in spatial mode, polarization, and optical frequency.
In practice, imperfect matching may reduce the interference contrast and limit the achievable background suppression.
The Jones matrices describing polarization rotation, an ideal quarter-wave plate (QWP), and EO-induced retardation are given by
\begin{gather}
    \bm{R}(\theta)=
    \begin{pmatrix}
        \cos\theta & -\sin\theta \\
        \sin\theta & \cos\theta
    \end{pmatrix}, \\
    \bm{QWP}=
    \begin{pmatrix}
        \exp\left(-\frac{i\pi}{4}\right) & 0 \\
        0 & \exp\left(\frac{i\pi}{4}\right)
    \end{pmatrix}, \\
    \bm{EO}(\Delta\phi)=
    \begin{pmatrix}
        \exp\left(-\frac{i\Delta\phi}{2}\right) & 0 \\
        0 & \exp\left(\frac{i\Delta\phi}{2}\right)
    \end{pmatrix},
\end{gather}
where $\theta$ denotes the rotation angle in the transverse polarization plane and $\Delta\phi$ denotes the signed EO-induced phase retardation. 
The QWP and EO matrices are expressed in their respective principal-axis bases.

We consider a (110)-cut EO crystal belonging to the $\bar{4}3m$ point group, with the probe incident normal to the crystal surface and the THz field polarized along the crystallographic $[\bar{1}10]$ axis. 
In this geometry, the EO principal axes are oriented at $\pm45^\circ$ relative to the $[\bar{1}10]$ axis. 
In the uniform-field approximation, the EO-induced retardation is given by \cite{Casalbuoni2008}
\begin{equation}
    \Delta\phi =
    \frac{2\pi n_0^3 d}{\lambda}r_{41}E_{\mathrm{THz}},
\end{equation}
where $E_{\mathrm{THz}}$ denotes the signed electric field inside the crystal along the $[\bar{1}10]$ axis. 
Here, $n_0$, $d$, $r_{41}$, and $\lambda$ denote the unperturbed refractive index at the probe wavelength, the crystal thickness, the EO coefficient, and the probe wavelength in vacuum, respectively. 
The angle $\alpha$ is defined as the angle between the incident probe polarization and the crystallographic $[\bar{1}10]$ axis, as illustrated in Fig.~\ref{Figure2}. 
To vary $\alpha$, the crystal and the THz polarization are rotated together about the probe propagation direction, keeping the THz field aligned with the $[\bar{1}10]$ axis. 
Thus, varying $\alpha$ changes the orientation of the probe polarization relative to the EO principal axes without changing $\Delta\phi$ at a fixed internal field.
Although $\Delta\phi$ itself remains unchanged, its manifestation in the probe polarization depends on $\alpha$: the probe becomes elliptically polarized after the EO crystal at $\alpha=0\tcdegree$, whereas it remains linearly polarized at $\alpha=45\tcdegree$.

In the numerical calculations, we consider a Ti:sapphire laser as a probe with a central wavelength of $\lambda=800~\mathrm{nm}$ and a (110)-cut ZnTe crystal. 
The probe is treated as monochromatic at this wavelength, and its finite spectral bandwidth is neglected. 
Unless otherwise stated, we use $n_0=2.85$~\cite{Marple1964} and $d=1.0~\mathrm{mm}$.
Following previous numerical studies of EOS with an $800~\mathrm{nm}$ probe, we adopt a constant EO coefficient of $r_{41}=4.0~\mathrm{pm/V}$ and neglect its frequency dependence~\cite{Casalbuoni2008}. 
The following analysis evaluates the optical readout for a prescribed $\Delta\phi$ and does not include the frequency-dependent THz transmission and propagation response of the crystal.
The signal maps are calculated for an internal THz field of $E_{\mathrm{THz}}=-3000~\mathrm{V/m}$, corresponding to $\Delta\phi\simeq-2.2\times10^{-3}~\mathrm{rad}$ for the crystal parameters above.

We denote the probe field incident on the EO crystal by $\bm{E}_{\mathrm{probe}}$ and the EO-modulated field entering the recombination beam splitter by $\bm{E}_{\mathrm{main}}$. 
The crystal is modeled as a lossless retarder, with Fresnel reflection losses at its surfaces neglected. 
The incident probe field, the EO-modulated probe field, and the unmodulated reference field $\bm{E}_{\mathrm{ref}}$ are written as
\begin{equation}
    \bm{E}_{\mathrm{probe}}=\sqrt{2}E_0
    \begin{pmatrix}
        \sqrt{\frac{1}{2}+\delta} \\
        0
    \end{pmatrix}, \\
\end{equation}
\begin{equation}
    \bm{E}_{\mathrm{main}}
    =\bm{R}\bigl(-(\alpha+45^\circ)\bigr)\cdot
    \bm{EO}(\Delta\phi)\cdot
    \bm{R}(\alpha+45^\circ)\cdot
    \bm{E}_{\mathrm{probe}},
\end{equation}
\begin{equation}
    \bm{E}_{\mathrm{ref}}=\sqrt{2}E_0
    \begin{pmatrix}
        \sqrt{\frac{1}{2}-\delta}\exp(-i\varepsilon) \\
        0
    \end{pmatrix}.
\end{equation}
Here, $E_0$ is a real normalization amplitude satisfying $\lVert\bm{E}_{\mathrm{probe}}\rVert^2+\lVert\bm{E}_{\mathrm{ref}}\rVert^2=2E_0^2$. 
The amplitude imbalance $\delta$ corresponds to the power fractions
$P_{\mathrm{main}}/P_{\mathrm{in}}=1/2+\delta$ and
$P_{\mathrm{ref}}/P_{\mathrm{in}}=1/2-\delta$, where
$P_{\mathrm{in}}=P_{\mathrm{main}}+P_{\mathrm{ref}}$ is the total
probe-plus-reference power incident on the recombination beam splitter.
The phase imbalance $\varepsilon$ is measured from the destructive-interference phase at the selected output port in the absence of EO modulation. 
The parameter ranges are $-1/2\leq\delta\leq1/2$ and $-\pi\leq\varepsilon<\pi$. 
Setting $\delta=\varepsilon=0$ gives complete cancellation of the field-independent background at this port.
We define the electric field immediately after the recombination non-polarizing beam splitter (NPBS) as
\begin{equation}
    \bm{E}_{\mathrm{out}}
    \equiv
    \frac{1}{\sqrt{2}}
    \left(
    \bm{E}_{\mathrm{main}}-\bm{E}_{\mathrm{ref}}
    \right).
    \label{eq:output_field_definition}
\end{equation}
The electric field incident on the detector is then expressed as
\begin{equation}
    \bm{E}_{\mathrm{det}}
    =
    \bm{M}\cdot\bm{E}_{\mathrm{out}},
    \label{eq:detector_field}
\end{equation}
where $\bm{M}$ denotes the Jones matrix representing the polarization-analysis optics between the recombination NPBS and the detector.
In the absence of polarization-analysis optics, $\bm{M}=\bm{I}$, and hence $\bm{E}_{\mathrm{det}}=\bm{E}_{\mathrm{out}}$.
The corresponding optical intensity at the detector is
\begin{equation}
    \begin{aligned}
    I_{\mathrm{det}}
    &=
    c\varepsilon_0\bm{E}_{\mathrm{det}}^{\dagger}\bm{E}_{\mathrm{det}}/2
    \\
    &=
    c\varepsilon_0
    \left(
    E_{\mathrm{det},x}^{*}E_{\mathrm{det},x}
    +E_{\mathrm{det},y}^{*}E_{\mathrm{det},y}
    \right)/2,
    \end{aligned}
    \label{eq:detector_intensity}
\end{equation}
where $c$ and $\varepsilon_0$ denote the speed of light in vacuum and the vacuum permittivity, respectively.
The $z$-axis is defined along the probe propagation direction, while the $x$- and $y$-axes span the transverse plane.
For all detection configurations, we use the following common definitions:
\begin{subequations}\label{eq:common-definitions}
\begin{align}
I(\Delta\phi)
&\equiv
I^{(0)}+I^{(1)}+I^{(2)}+I^{(3)}+\cdots,
\label{eq:response-expansion}
\\
I_{\mathrm{sig}}(\Delta\phi)
&\equiv
I(\Delta\phi)-I(0),
\label{eq:signal-definition}
\end{align}
\end{subequations}
where $I^{(n)}$ denotes the contribution proportional to $(\Delta\phi)^n$, with $I^{(0)}=I(0)$.
For a single-channel configuration,
\begin{equation}
    I(\Delta\phi)\equiv I_{\mathrm{det}}(\Delta\phi),
\end{equation}
whereas for a differential configuration,
\begin{equation}
    \begin{aligned}
    I(\Delta\phi)
    &\equiv
    I_{\mathrm{diff}}(\Delta\phi),
    \\
    I_{\mathrm{diff}}(\Delta\phi)
    &\equiv
    I_{\mathrm{det},1}(\Delta\phi)-I_{\mathrm{det},2}(\Delta\phi).
    \end{aligned}
    \label{eq:detected-response}
\end{equation}
For normalization, we define
\begin{equation}
    I_{\mathrm{tot}}(\Delta\phi)
    \equiv
    \max_{1\leq j\leq j_{\mathrm{max}}}
    \left\{
    I_{\mathrm{det},j}(\Delta\phi)
    \right\},
    \label{eq:normalization-intensity}
\end{equation}
where $j$ labels the individual photodiodes, with $j_{\mathrm{max}}=1$ ($I_{\mathrm{det},1}\equiv I_{\mathrm{det}}$) for a single-channel configuration and $j_{\mathrm{max}}=2$ for a two-channel configuration.

The normalization $I_{\mathrm{sig}}/I_{\mathrm{tot}}$ expresses the EO-induced intensity change relative to the optical intensity incident on an individual detector, enabling direct comparison of the relative EO modulation among different detection configurations.
The normalization intensity
$I_{\mathrm{tot}}$ is evaluated from the exact, unexpanded
detector outputs.

\subsection{Detection without polarization-analysis optics}
We derive the detected intensities for the two crystal orientations $\alpha=0\tcdegree$ and $\alpha=45\tcdegree$ shown in Fig.~\ref{Figure2}, with no polarization-analysis optics between recombination NPBS and the detector.

\subsubsection{Configuration with \texorpdfstring{$\alpha=0\tcdegree$}{alpha=0 degrees}}
For $\alpha=0\tcdegree$, as shown in Fig.~\ref{Figure2}(a), the field at the nominally dark output port of the recombination NPBS is
\begin{equation}
    \begin{aligned}
        \bm{E}_{\mathrm{out}}
        &=
        \frac{1}{\sqrt{2}}
        \Bigl[
        \bm{R}(-45\tcdegree)\cdot
        \bm{EO}(\Delta\phi)\cdot
        \bm{R}(45\tcdegree)\cdot
        \bm{E}_{\mathrm{probe}}
        -\bm{E}_{\mathrm{ref}}
        \Bigr] \\
        &=E_0
        \begin{pmatrix}
            \sqrt{\frac{1}{2}+\delta}\cos\left(\frac{\Delta\phi}{2}\right)
            -\sqrt{\frac{1}{2}-\delta}\exp(-i\varepsilon) \\
            i\sqrt{\frac{1}{2}+\delta}\sin\left(\frac{\Delta\phi}{2}\right)
        \end{pmatrix}.
    \end{aligned}
    \label{eq:alpha0_output_field}
\end{equation}
Since no polarization-analysis optics are placed between the recombination NPBS and the detector, $\bm{E}_{\mathrm{det}}=\bm{E}_{\mathrm{out}}$.
The corresponding optical intensity at the detector is
\begin{equation}
    \begin{aligned}
        I_{\mathrm{det}}
        &=
        \frac{c\varepsilon_0E_0^2}{2}
        \left(
        1-\sqrt{1-4\delta^2}
        \cos\varepsilon
        \cos\left(\frac{\Delta\phi}{2}\right)
        \right) \\
        &=
        \frac{c\varepsilon_0E_0^2}{2}
        \Biggl[
        1-\sqrt{1-4\delta^2}\cos\varepsilon \\
        &\qquad+
        \frac{\sqrt{1-4\delta^2}\cos\varepsilon}{8}
        (\Delta\phi)^2
        +\mathcal{O}\bigl((\Delta\phi)^4\bigr)
        \Biggr].
    \end{aligned}
    \label{eq:alpha0_output_intensity}
\end{equation}

As shown in Fig.~\ref{Figure3}(a), the normalized signal $I_{\mathrm{sig}}/I_{\mathrm{tot}}$ increases toward the background-cancellation point $\delta=\varepsilon=0$, reaching unity at the exact cancellation for $\Delta\phi\neq0$.
However, the normalized first-order signal $I^{(1)}/I_{\mathrm{tot}}$ vanishes throughout the parameter range, as shown in Fig.~\ref{Figure3}(b), because $I_{\mathrm{det}}$ is an even function of $\Delta\phi$.

At $\delta=\varepsilon=0$, the field-independent $x$-polarized component is completely canceled.
The EO-induced $y$-polarized field is first order in $\Delta\phi$ but has no reference component of the same polarization with which to interfere.
Its contribution to the detected intensity is therefore second order, whereas the residual $x$-polarized intensity is fourth order.
Thus, in this configuration, complete background cancellation maximizes the normalized signal but yields a quadratic rather than linear weak-field response, and the detected intensity does not distinguish opposite THz-field polarities.

\subsubsection{Configuration with \texorpdfstring{$\alpha=45\tcdegree$}{alpha=45 degrees}}
\begin{figure}[!t]
\centering\includegraphics[width=8.2cm]{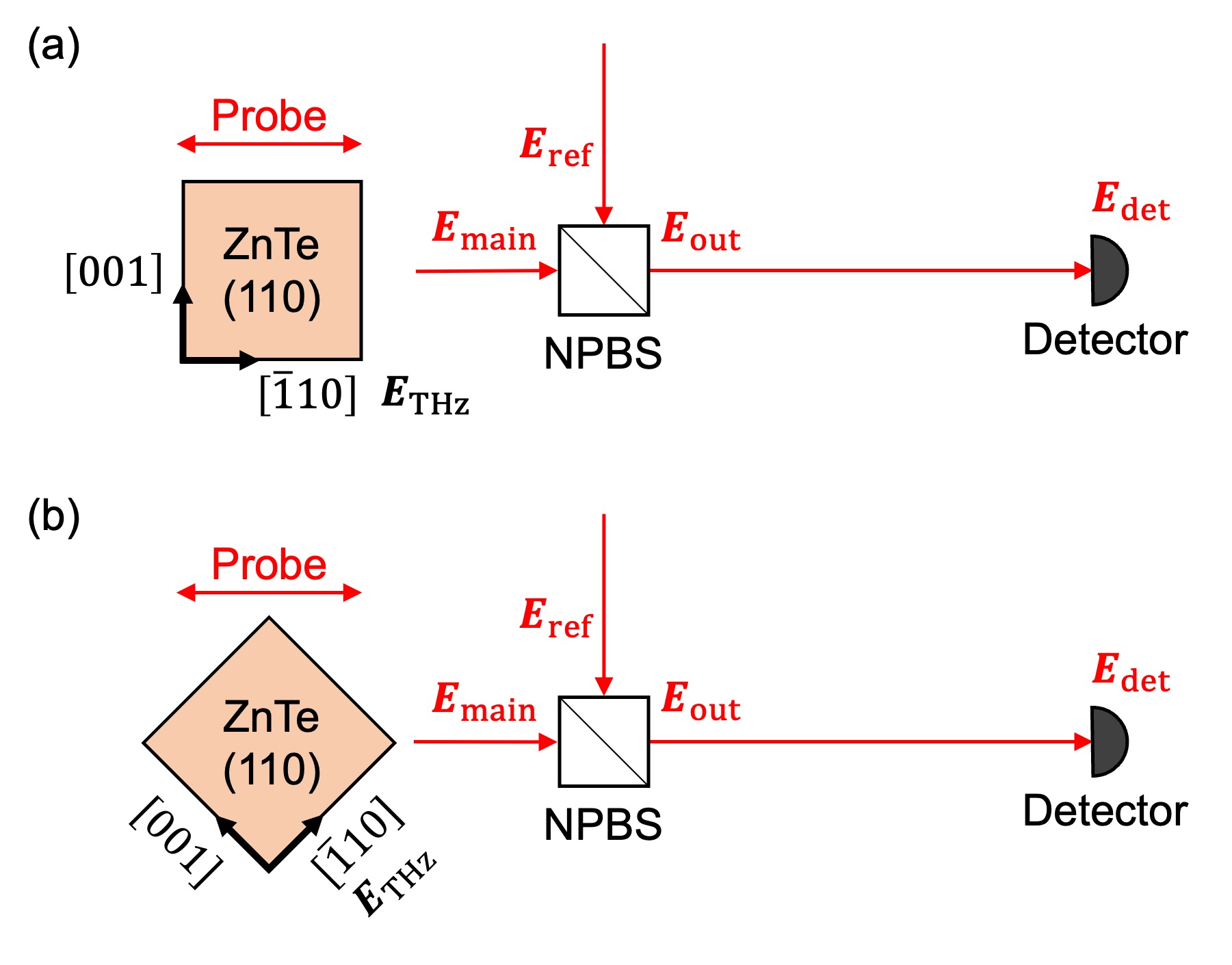}
\caption{
Schematics of background-suppressed electro-optic (EO) configurations without polarization-analysis optics in front of the detector.
The probe polarization is fixed, while the EO crystal and the THz polarization are rotated together so that their relative orientation remains unchanged:
(a) $\alpha=0\tcdegree$, where the crystallographic $[\bar{1}10]$ axis and the THz polarization are parallel to the probe polarization, and
(b) $\alpha=45\tcdegree$, where the crystallographic $[\bar{1}10]$ axis and the THz polarization are oriented at $45\tcdegree$ with respect to the probe polarization.
NPBS: non-polarizing beam splitter.
}
\label{Figure2}
\end{figure}

The configuration for $\alpha=45\tcdegree$ is shown in Fig.~\ref{Figure2}(b).
The field at the nominally dark output port is
\begin{equation}
    \begin{aligned}
        \bm{E}_{\mathrm{out}}
        &=
        \frac{1}{\sqrt{2}}
        \Bigl[
        \bm{R}(-90\tcdegree)\cdot
        \bm{EO}(\Delta\phi)\cdot
        \bm{R}(90\tcdegree)\cdot
        \bm{E}_{\mathrm{probe}}
        -\bm{E}_{\mathrm{ref}}
        \Bigr] \\
        &=E_0
        \begin{pmatrix}
            \sqrt{\frac{1}{2}+\delta}\exp\left(i\frac{\Delta\phi}{2}\right)
            -\sqrt{\frac{1}{2}-\delta}\exp(-i\varepsilon) \\
            0
        \end{pmatrix}.
    \end{aligned}
    \label{eq:alpha45_output_field}
\end{equation}

The corresponding detected intensity is
\begin{equation}
    \begin{aligned}
        I_{\mathrm{det}}
        &=
        \frac{c\varepsilon_0E_0^2}{2}
        \left[
        1-\sqrt{1-4\delta^2}
        \cos\left(\varepsilon+\frac{\Delta\phi}{2}\right)
        \right] \\
        &=
        \frac{c\varepsilon_0E_0^2}{2}
        \Biggl[
        1-\sqrt{1-4\delta^2}\cos\varepsilon \\
        &\qquad+
        \frac{\sqrt{1-4\delta^2}}{2}
        \sin\varepsilon\,\Delta\phi \\
        &\qquad+
        \frac{\sqrt{1-4\delta^2}}{8}
        \cos\varepsilon\,(\Delta\phi)^2
        +\mathcal{O}\bigl((\Delta\phi)^3\bigr)
        \Biggr].
    \end{aligned}
    \label{eq:alpha45_output_intensity}
\end{equation}

\begin{figure}[!t]
\centering\includegraphics[width=8.2cm]{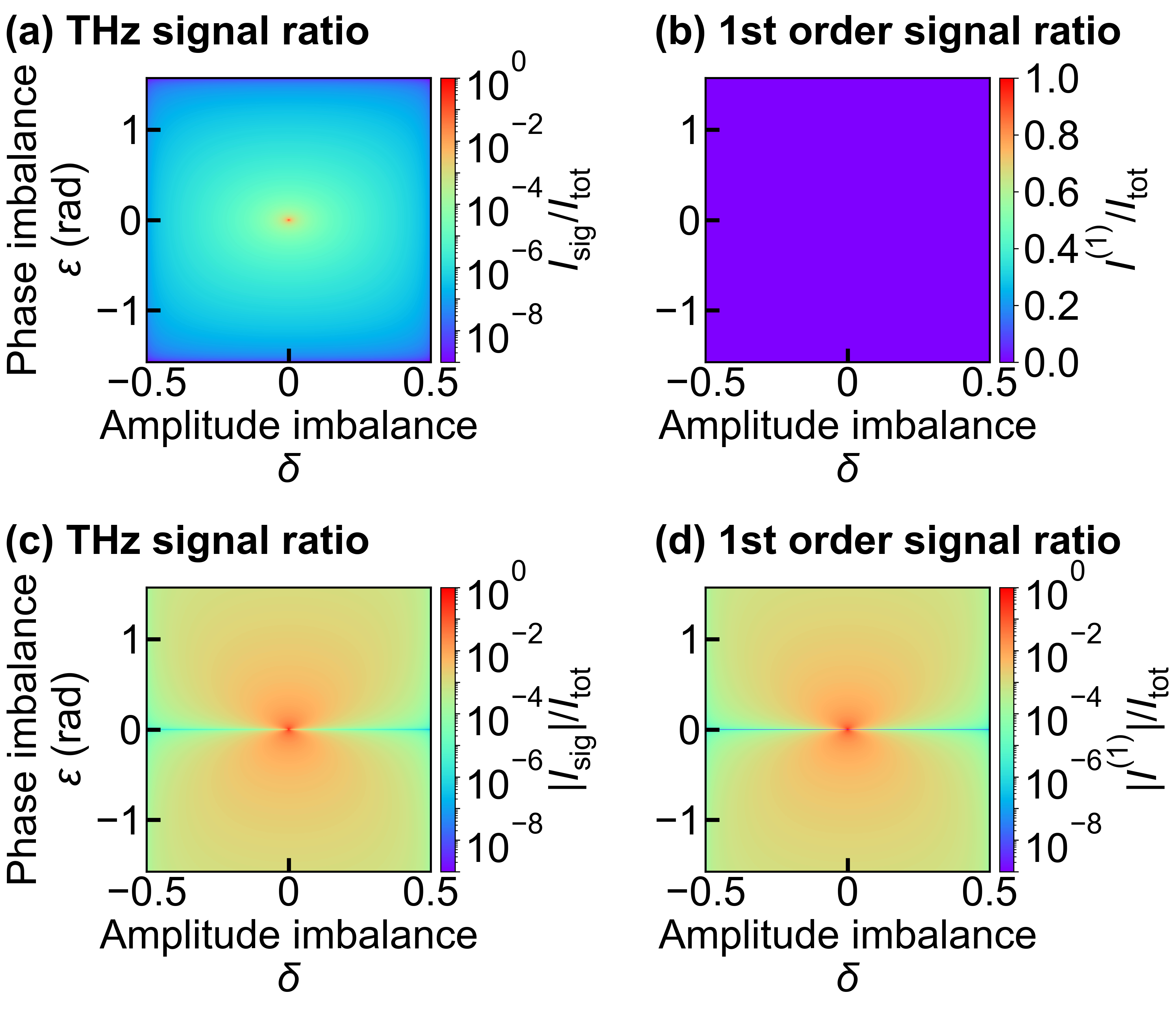}
\caption{Calculated two-dimensional color maps of the normalized THz-induced signal and first-order signal for $\alpha=0\tcdegree$ and $45\tcdegree$ as functions of the amplitude imbalance $\delta$ and phase imbalance $\varepsilon$. 
The upper row corresponds to $\alpha=0\tcdegree$: 
(a) $I_{\mathrm{sig}}/I_{\mathrm{tot}}$ and 
(b) $I^{(1)}/I_{\mathrm{tot}}$. 
The lower row corresponds to $\alpha=45\tcdegree$: 
(c) $|I_{\mathrm{sig}}|/I_{\mathrm{tot}}$ and 
(d) $|I^{(1)}|/I_{\mathrm{tot}}$. 
Panels (a), (c), and (d) are plotted on logarithmic color scales with a common lower bound of $10^{-9}$, whereas panel (b), which is identically zero over the entire parameter range, is plotted on a linear color scale.}
\label{Figure3}
\end{figure}

As shown in Fig.~\ref{Figure3}(c), the normalized signal magnitude
$|I_{\mathrm{sig}}|/I_{\mathrm{tot}}$ becomes large near $\delta=0$
at small but finite phase imbalance.
The absolute value is used to compare signal magnitudes irrespective of their signs.
Figure~\ref{Figure3}(d) shows a similar distribution for the normalized
first-order signal $|I^{(1)}|/I_{\mathrm{tot}}$, in contrast to the
identically zero first-order response in Fig.~\ref{Figure3}(b).
The close agreement between Figs.~\ref{Figure3}(c) and
\ref{Figure3}(d) away from the vicinity of $\varepsilon=0$
indicates that the response is predominantly first order in that
region under the weak-field conditions.

In this configuration, the probe polarization is aligned with one
of the EO principal axes, so the EO interaction introduces a phase
shift of $\Delta\phi/2$ without changing the polarization state.
Interference with the unmodulated reference beam converts this optical phase shift into a measurable intensity change without requiring polarization-analysis optics.
The first-order response is nonzero when $|\delta|<1/2$ and
$\sin\varepsilon\neq0$, and its sign depends on the phase
imbalance through
$I^{(1)} \propto \sin\varepsilon\,\Delta\phi$.
At $\varepsilon=0$, however, the first-order term vanishes, and the remaining EO signal arises from even-order contributions.
Thus, this configuration illustrates polarization-analysis-free detection of an EO-induced optical phase shift, although a finite interferometer phase bias is required to retain a linear weak-field response.

\begin{figure}[ht]
\centering\includegraphics[width=8.2cm]{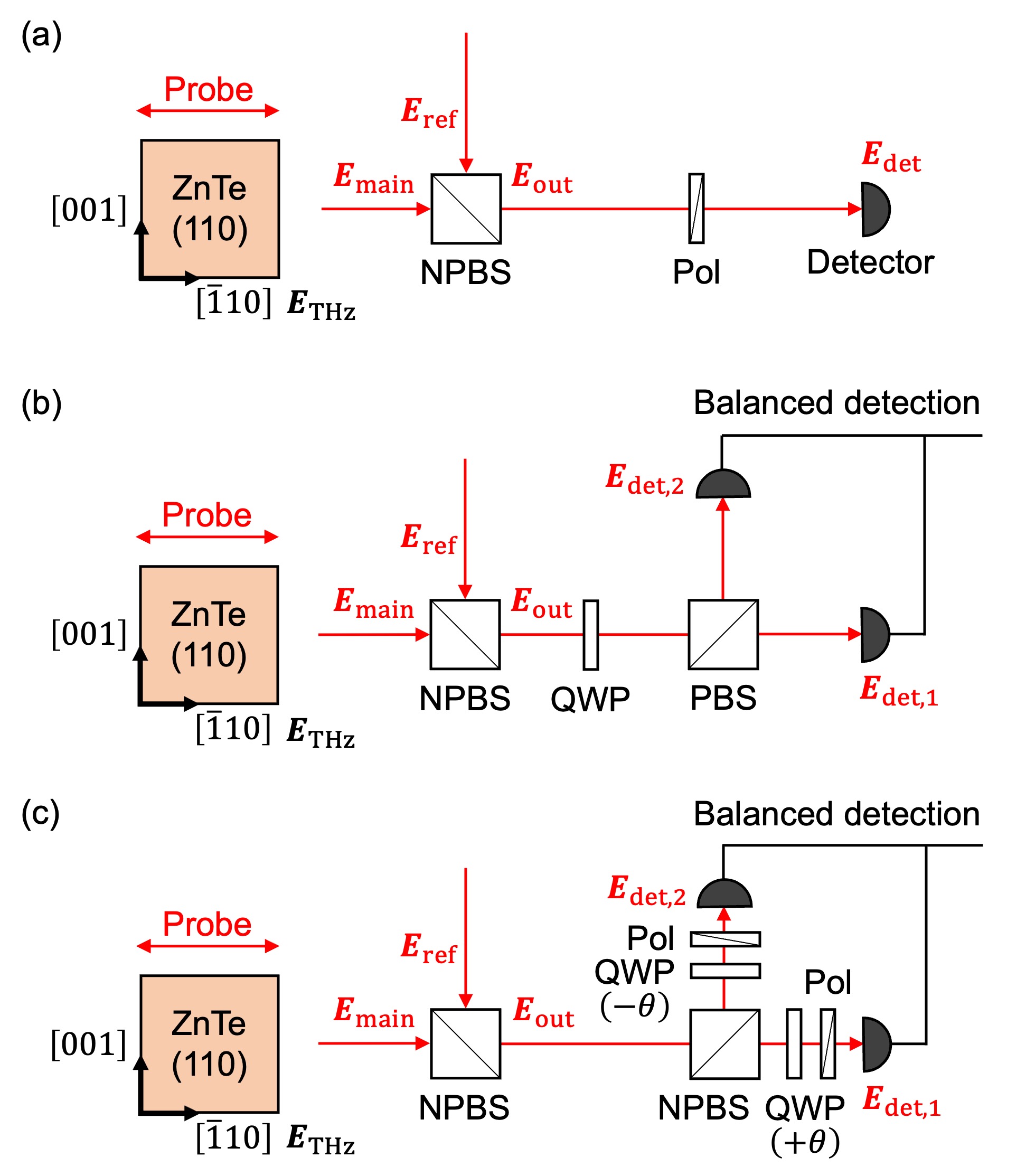}
\caption{Schematics of background-suppressed electro-optic (EO) configurations with polarization-analysis optics: 
(a) crossed-Nicols configuration, 
(b) balanced-detection configuration, and 
(c) phase-offset configuration. 
In all three configurations, the crystal orientation is $\alpha=0\tcdegree$. 
NPBS: non-polarizing beam splitter;
Pol: polarizer;
QWP: quarter-wave plate; 
PBS: polarizing beam splitter.}
\label{Figure4}
\end{figure}

\subsection{Detection with polarization-analysis optics}
The $\alpha=0\tcdegree$ orientation converts the
EO-induced retardation into a polarization change, which can be
analyzed with additional polarization optics.
In the following analysis, we take the output field for $\alpha=0\tcdegree$ (Eq.~(\ref{eq:alpha0_output_field})) as the input field to the polarization-analysis optics.
We consider three representative EO readout schemes with polarization analysis: the crossed-Nicols, balanced-detection, and phase-offset configurations \cite{Brunner2014,Pan2014}.

\subsubsection{Crossed-Nicols configuration}
The crossed-Nicols configuration is illustrated in Fig.~\ref{Figure4}(a).
The analyzer is aligned to transmit the $y$-polarized component and suppress the $x$-polarized component.
Its finite extinction is modeled by the Jones matrix
\begin{equation}
    \bm{POL}
    =
        \begin{pmatrix}
        10^{-\mathrm{OD}/2} & 0 \\
        0 & 1
    \end{pmatrix}, \\
\end{equation}
where $\mathrm{OD}$ denotes the optical density for the rejected $x$-polarized component.
Unit intensity transmission is assumed for the $y$-polarized component.
We use $\mathrm{OD}=3$ in the numerical calculations for configurations employing this polarizer.

The field transmitted by the analyzer is
\begin{equation}
    \bm{E}_{\mathrm{det}}
    =
    \bm{POL}\cdot\bm{E}_{\mathrm{out}}.
    \label{eq:crossed_nicols_field}
\end{equation}
The corresponding detected intensity is
\begin{equation}
    \begin{aligned}
        I_{\mathrm{det}}
        &=
        \frac{c\varepsilon_0 E_0^2}{2}
        \Biggl[
        10^{-\mathrm{OD}}
        \Biggl\{
        \left(\frac{1}{2}+\delta\right)
        \cos^2\left(\frac{\Delta\phi}{2}\right)
        \\
        &\qquad
        +\left(\frac{1}{2}-\delta\right)
        -\sqrt{1-4\delta^2}
        \cos\left(\frac{\Delta\phi}{2}\right)
        \cos\varepsilon
        \Biggr\}
        \\
        &\qquad
        +\left(\frac{1}{2}+\delta\right)
        \sin^2\left(\frac{\Delta\phi}{2}\right)
        \Biggr]
        \\
        &=
        \frac{c\varepsilon_0 E_0^2}{2}
        \Biggl[
        10^{-\mathrm{OD}}
        \left(
        1-\sqrt{1-4\delta^2}\cos\varepsilon
        \right)
        \\
        &\qquad
        +\frac{1}{8}
        \Biggl\{
        (1+2\delta)
        \left(1-10^{-\mathrm{OD}}\right)
        \\
        &\qquad\qquad
        +10^{-\mathrm{OD}}
        \sqrt{1-4\delta^2}\cos\varepsilon
        \Biggr\}
        (\Delta\phi)^2
        \\
        &\qquad
        +\mathcal{O}\bigl((\Delta\phi)^4\bigr)
        \Biggr].
    \end{aligned}
    \label{eq:crossed_nicols_intensity}
\end{equation}

\begin{figure}[!b]
\centering\includegraphics[width=8.2cm]{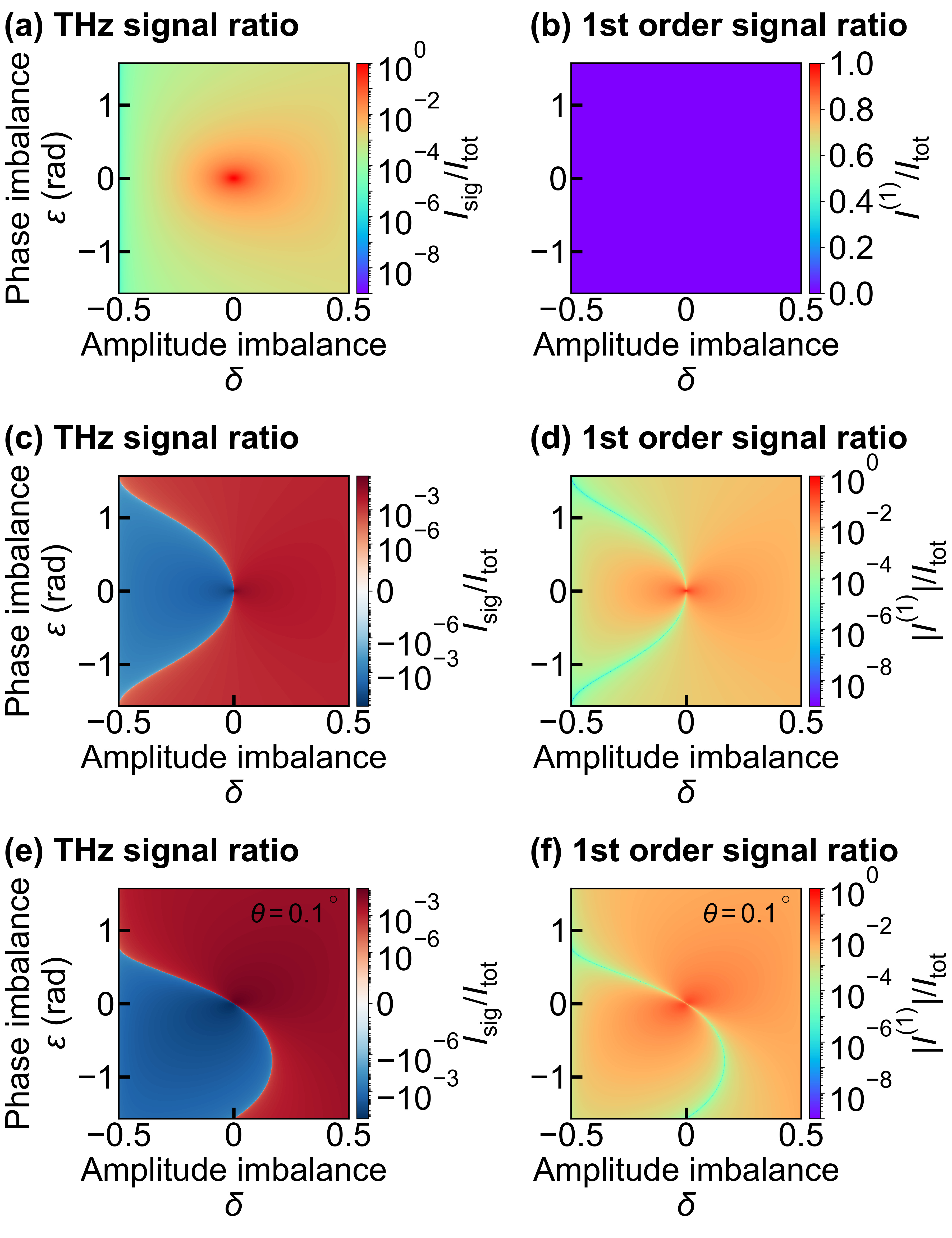}
\caption{Calculated two-dimensional color maps of the normalized THz-induced signal and first-order signal for the crossed-Nicols, balanced-detection, and phase-offset configurations as functions of the amplitude imbalance $\delta$ and phase imbalance $\varepsilon$. 
The upper row corresponds to the crossed-Nicols configuration with $\mathrm{OD}=3$: 
(a) $I_{\mathrm{sig}}/I_{\mathrm{tot}}$ and 
(b) $I^{(1)}/I_{\mathrm{tot}}$. 
The middle row corresponds to the balanced-detection configuration with a PBS extinction ratio of $\cot^2\chi=10^{3}$:
(c) $I_{\mathrm{sig}}/I_{\mathrm{tot}}$ and 
(d) $|I^{(1)}|/I_{\mathrm{tot}}$. 
The lower row corresponds to the phase-offset configuration with $\theta=0.1\tcdegree$ and $\mathrm{OD}=3$: 
(e) $I_{\mathrm{sig}}/I_{\mathrm{tot}}$ and 
(f) $|I^{(1)}|/I_{\mathrm{tot}}$. 
Panels (a), (d), and (f) are plotted on logarithmic color scales with a lower bound of $10^{-9}$. 
Panels (c) and (e) are plotted on symmetric logarithmic color scales with a linear threshold of $10^{-9}$ to represent both positive and negative values, whereas panel (b), which is identically zero over the entire parameter range, is plotted on a linear color scale.}
\label{Figure5}
\end{figure}

As shown in Fig.~\ref{Figure5}(a), the normalized signal
$I_{\mathrm{sig}}/I_{\mathrm{tot}}$ becomes large near
$\delta=\varepsilon=0$ for the finite polarizer extinction
($\mathrm{OD}=3$) considered here.
The zero-field background is proportional to
$10^{-\mathrm{OD}}\left(1-\sqrt{1-4\delta^2}\cos\varepsilon\right)$,
showing that polarization filtering and interferometric suppression
act multiplicatively.
Destructive interference can therefore further suppress the coherent
background transmitted through the finite-extinction polarizer.
At $\delta=\varepsilon=0$, the zero-field background vanishes,
giving $I_{\mathrm{sig}}/I_{\mathrm{tot}}=1$ for the nonzero
retardation considered here.

However, Fig.~\ref{Figure5}(b) shows that the normalized first-order
signal $I^{(1)}/I_{\mathrm{tot}}$ is zero throughout the parameter range.
This follows from the even dependence of $I_{\mathrm{det}}$ on
$\Delta\phi$, which gives $I^{(1)}=0$ for all $\delta$ and
$\varepsilon$ and
$I_{\mathrm{sig}}=I^{(2)}+\mathcal{O}\bigl((\Delta\phi)^4\bigr)$.
Thus, combining polarization filtering with interferometric
background suppression increases the normalized EO signal near
the cancellation point but does not produce a linear weak-field
response, and the detected intensity does not distinguish opposite
THz-field polarities.

\subsubsection{Balanced-detection configuration}
The balanced-detection configuration is illustrated in Fig.~\ref{Figure4}(b).
After recombination, the field passes through a QWP oriented at $45\tcdegree$ relative to the horizontal polarization axis.
A polarizing beam splitter (PBS) then separates the horizontal and vertical polarization components for detection with two photodiodes.
The PBS is modeled by the Jones matrices
\begin{equation}
    \bm{PBS}_1
    =
    \begin{pmatrix}
        \cos\chi & 0 \\
        0 & \sin\chi
    \end{pmatrix},
    \qquad
    \bm{PBS}_2
    =
    \begin{pmatrix}
        \sin\chi & 0 \\
        0 & \cos\chi
    \end{pmatrix},
    \label{eq:pbs_matrices}
\end{equation}
where ports 1 and 2 correspond predominantly to the horizontal and vertical polarization components, respectively.
The angle $\chi$ describes the leakage of the rejected polarization component into each port, giving an extinction ratio of $\cot^2\chi$ at both ports.
Because $\cos^2\chi+\sin^2\chi=1$, the PBS conserves the total optical power; $\chi=0$ corresponds to an ideal PBS.
The electric fields at ports 1 and 2 are
\begin{equation}
    \begin{aligned}
        \bm{E}_{\mathrm{det},1}
        &=
        \bm{PBS}_1\cdot
        \bm{R}(-45\tcdegree)\cdot
        \bm{QWP}\cdot
        \bm{R}(45\tcdegree)\cdot
        \bm{E}_{\mathrm{out}}
        \\
        &=
        \begin{pmatrix}
            E_{\mathrm{H}}\cos\chi \\
            E_{\mathrm{V}}\sin\chi
        \end{pmatrix},
        \\
        \bm{E}_{\mathrm{det},2}
        &=
        \bm{PBS}_2\cdot
        \bm{R}(-45\tcdegree)\cdot
        \bm{QWP}\cdot
        \bm{R}(45\tcdegree)\cdot
        \bm{E}_{\mathrm{out}}
        \\
        &=
        \begin{pmatrix}
            E_{\mathrm{H}}\sin\chi \\
            E_{\mathrm{V}}\cos\chi
        \end{pmatrix},
    \end{aligned}
    \label{eq:c2_output_field}
\end{equation}
where $E_{\mathrm{H}}$ and $E_{\mathrm{V}}$ denote the horizontal ($x$) and vertical ($y$) components of $\bm{R}(-45\tcdegree)\cdot\bm{QWP}\cdot\bm{R}(45\tcdegree)\cdot\bm{E}_{\mathrm{out}}$, respectively:
\begin{equation}
    \begin{aligned}
        E_{\mathrm{H}}
        &=
        \frac{E_0}{\sqrt{2}}
        \Biggl[
        \sqrt{\frac{1}{2}+\delta}
        \left\{
        \cos\left(\frac{\Delta\phi}{2}\right)
        -
        \sin\left(\frac{\Delta\phi}{2}\right)
        \right\}
        \\
        &\qquad -
        \sqrt{\frac{1}{2}-\delta}\,
        \exp(-i\varepsilon)
        \Biggr],
        \\
        E_{\mathrm{V}}
        &=
        \frac{iE_0}{\sqrt{2}}
        \Biggl[
        \sqrt{\frac{1}{2}+\delta}
        \left\{
        \cos\left(\frac{\Delta\phi}{2}\right)
        +
        \sin\left(\frac{\Delta\phi}{2}\right)
        \right\}
        \\
        &\qquad -
        \sqrt{\frac{1}{2}-\delta}\,
        \exp(-i\varepsilon)
        \Biggr].
    \end{aligned}
    \label{eq:balanced_field_components}
\end{equation}
The intensities incident on the two photodiodes are
\begin{equation}
    I_{\mathrm{det},j}
    =
    \frac{c\varepsilon_0}{2}
    \lVert\bm{E}_{\mathrm{det},j}\rVert^2,
\end{equation}
which give $I_{\mathrm{det},1}=I_{\mathrm{H}}\cos^2\chi+I_{\mathrm{V}}\sin^2\chi$ and $I_{\mathrm{det},2}=I_{\mathrm{H}}\sin^2\chi+I_{\mathrm{V}}\cos^2\chi$, where $I_{\mathrm{H}}=c\varepsilon_0|E_{\mathrm{H}}|^2/2$ and $I_{\mathrm{V}}=c\varepsilon_0|E_{\mathrm{V}}|^2/2$.
The differential output is
\begin{equation}
    \begin{aligned}
        I_{\mathrm{diff}}
        &=
        \cos 2\chi
        \left(I_{\mathrm{H}}-I_{\mathrm{V}}\right)
        \\
        &=
        \frac{c\varepsilon_0E_0^2}{2}\cos 2\chi
        \Biggl[
        \sqrt{1-4\delta^2}\cos\varepsilon
        \sin\left(\frac{\Delta\phi}{2}\right)
        \\
        &\qquad
        -\frac{1+2\delta}{2}\sin\Delta\phi
        \Biggr]
        \\
        &=
        \frac{c\varepsilon_0E_0^2}{2}\cos 2\chi
        \Biggl[
        \frac{
        \sqrt{1-4\delta^2}\cos\varepsilon-(1+2\delta)
        }{2}\Delta\phi
        \\
        &\qquad
        +
        \frac{
        4(1+2\delta)-\sqrt{1-4\delta^2}\cos\varepsilon
        }{48}
        (\Delta\phi)^3
        \\
        &\qquad
        +\mathcal{O}\bigl((\Delta\phi)^5\bigr)
        \Biggr].
    \end{aligned}
    \label{eq:balanced_difference}
\end{equation}
The finite extinction of the PBS thus only rescales the differential output by $\cos 2\chi$ to all orders in $\Delta\phi$, whereas the zero-field intensity on each photodiode, $I_{\mathrm{det},1}(0)=I_{\mathrm{det},2}(0)=I_{\mathrm{H}}(0)$, is independent of $\chi$ because $I_{\mathrm{H}}(0)=I_{\mathrm{V}}(0)$.
For a high-extinction PBS, $\cos 2\chi\simeq1-2\chi^2$ is close to unity.
The numerical results for balanced detection in this paper are calculated for $\cot^2\chi=10^{3}$, i.e., the same extinction ratio as that of the polarizer with $\mathrm{OD}=3$, for which $\cos 2\chi\simeq0.998$.

As shown in Fig.~\ref{Figure5}(c), the normalized differential signal
$I_{\mathrm{sig}}/I_{\mathrm{tot}}$ exhibits both positive and
negative regions for $\delta<0$, whereas its polarity is maintained
as $\varepsilon$ varies on the $\delta>0$ branch in the
linear-response regime.
Figure~\ref{Figure5}(d) shows that the normalized first-order signal
$|I^{(1)}|/I_{\mathrm{tot}}$ is nonzero at $\varepsilon=0$ for
$\delta>0$ and becomes large at small but finite amplitude imbalance.
Thus, interferometric background suppression can yield a large
normalized linear signal.
For $\delta > 0$,
$\sqrt{1-4\delta^2}\cos\varepsilon < (1+2\delta)$ for
all $\varepsilon$, so the first-order coefficient in Eq.~(\ref{eq:balanced_difference}),
which contains the positive factor $\cos 2\chi$, remains negative.
Phase fluctuations therefore do not reverse the first-order
signal on this branch.

At $\delta=\varepsilon=0$, however, the first-order coefficient
vanishes and the leading differential signal is cubic in $\Delta\phi$.
At zero phase imbalance, a finite positive amplitude imbalance
leaves a residual background field that is mixed with the
EO-induced orthogonally polarized field by the QWP--PBS analyzer,
providing the first-order response.

The odd symmetry of Eq.~(\ref{eq:balanced_difference}) also
provides an advantage over single-channel EO sampling operated
at a single optical bias near the crossed-polarizer extinction
point, where a quadratic field-dependent term can distort the
measured waveform \cite{Jiang1999}.
In the present configuration, subtraction of the two simultaneously
detected polarization components cancels all even-order terms,
eliminating this source of waveform distortion without optical-bias
reversal or nonlinear post-processing.

\subsubsection{Phase-offset configuration}
The phase-offset configuration is illustrated in Fig.~\ref{Figure4}(c).
After recombination, an additional ideal 50:50 beam splitter divides the field into two detection channels.
The two channels use QWPs with opposite rotation angles ($\pm\theta$) and identical finite-extinction polarizers.
The corresponding output fields are
\begin{equation}
    \bm{E}_{\mathrm{det},j}
    =
    \frac{1}{\sqrt{2}}\cdot
    \bm{POL}\cdot
    \bm{R}(\pm\theta)\cdot
    \bm{QWP}\cdot
    \bm{R}(\mp\theta)\cdot
    \bm{E}_{\mathrm{out}}.
    \label{eq:phase_offset_field}
\end{equation}
Here, the subscripts $j$ ($j=1,2$) label the two detection channels.
The upper and lower signs correspond to $j=1$ and $2$, respectively.
The factor $1/\sqrt{2}$ accounts for the additional beam splitter.
The differential output is
\begin{equation}
    \begin{aligned}
        I_{\mathrm{diff}}
        &=
        -\frac{c\varepsilon_0E_0^2}{4}
        \left(1-10^{-\mathrm{OD}}\right)
        \sin\left(\frac{\Delta\phi}{2}\right)
        \sin 2\theta
        \\
        &\quad\times
        \Biggl[
        (1+2\delta)
        \cos\left(\frac{\Delta\phi}{2}\right)
        \\
        &\qquad\qquad
        -\sqrt{1-4\delta^2}
        \bigl(\cos\varepsilon-\cos 2\theta\sin\varepsilon\bigr)
        \Biggr]
        \\
        &=
        \frac{c\varepsilon_0E_0^2}{2}
        \left(1-10^{-\mathrm{OD}}\right)
        \sin 2\theta
        \\
        &\quad\times
        \Biggl[
        -\frac{1}{4}
        \Biggl\{
        (1+2\delta)
        \\
        &\qquad\qquad
        -\sqrt{1-4\delta^2}
        \bigl(\cos\varepsilon-\cos 2\theta\sin\varepsilon\bigr)
        \Biggr\}\Delta\phi
        \\
        &\qquad
        +\frac{1}{96}
        \Biggl\{
        4(1+2\delta)
        \\
        &\qquad\qquad
        -\sqrt{1-4\delta^2}
        \bigl(\cos\varepsilon-\cos 2\theta\sin\varepsilon\bigr)
        \Biggr\}(\Delta\phi)^3
        \\
        &\qquad
        +\mathcal{O}\bigl((\Delta\phi)^5\bigr)
        \Biggr].
    \end{aligned}
    \label{eq:phase_offset_difference}
\end{equation}

Figures~\ref{Figure5}(e) and \ref{Figure5}(f) show the normalized
differential signal $I_{\mathrm{sig}}/I_{\mathrm{tot}}$ and
the normalized first-order signal magnitude
$|I^{(1)}|/I_{\mathrm{tot}}$, respectively, for
$\theta=0.1\tcdegree$ and $\mathrm{OD}=3$.
The polarity of the linear signal is determined by the sign of the first-order coefficient in Eq.~(\ref{eq:phase_offset_difference}), which is proportional to $(1+2\delta)-\sqrt{1-4\delta^2}\,(\cos\varepsilon-\cos2\theta\sin\varepsilon)$.
Because of the term proportional to $\cos2\theta\sin\varepsilon$, which is linear in $\varepsilon$, this coefficient vanishes for the small QWP rotation used here along a curve that passes through $\delta=\varepsilon=0$ and extends into the $\delta>0$ region.
As shown in Fig.~\ref{Figure5}(e), this curve separates regions of opposite signal polarity, and Fig.~\ref{Figure5}(f) shows that $|I^{(1)}|/I_{\mathrm{tot}}$ is suppressed along the same curve.
A nonzero first-order response is obtained at $\varepsilon=0$ for finite positive amplitude imbalance, whereas at $\delta=\varepsilon=0$ the first-order coefficient vanishes and the leading differential signal is cubic.

\section{Optimization of background-suppressed EO sampling}
\label{sec:optimization}

We now compare the detection configurations in terms of their
weak-field linearity and the robustness of the response
against interferometer phase fluctuations.

Without polarization-analysis optics at $\alpha=0\tcdegree$
(Fig.~\ref{Figure2}(a)), and in the crossed-Nicols configuration
(Fig.~\ref{Figure4}(a)), the detected intensity is an even
function of $\Delta\phi$ and therefore does not resolve the
THz-field polarity.
The configuration without polarization-analysis optics at
$\alpha=45\tcdegree$ (Fig.~\ref{Figure2}(b)) provides a first-order response when
$|\delta|<1/2$ and $\sin\varepsilon\neq0$, although the quadratic
contribution must remain small for faithful linear readout.
Both differential configurations cancel all even-order terms
and provide a linear weak-field response at operating points
with a nonzero first-order coefficient.
At the exact background-cancellation point, however, their
leading response is cubic rather than linear for the analyzer
settings considered here.

Interferometer phase fluctuations, mainly caused by mechanical
vibrations of the optical components, lead to variations in the
phase imbalance $\varepsilon$, corresponding
to vertical motion in the maps of Figs.~\ref{Figure3} and \ref{Figure5} at fixed $\delta$.
In EO sampling, the THz waveform is usually reconstructed from many probe pulses by scanning the probe delay.
Shot-to-shot fluctuations of $\varepsilon$ that change the first-order response coefficient therefore translate directly into amplitude noise and distortion of the measured waveform, and a polarity reversal is the extreme case in which this coefficient changes sign.
For the configuration without polarization-analysis optics at
$\alpha=45^\circ$ (Fig.~\ref{Figure2}(b)), the first-order response changes sign across
$\varepsilon=0$ because
$I^{(1)}\propto\sin\varepsilon\,\Delta\phi$.
In the balanced-detection configuration (Fig.~\ref{Figure4}(b)), by contrast, the
first-order coefficient retains its sign for all $\varepsilon$ on
the $\delta>0$ branch, as follows from Eq.~(\ref{eq:balanced_difference}).
For the phase-offset configuration, phase fluctuations can move the operating point across the polarity boundary in Fig.~\ref{Figure5}(e) and thereby reverse the signal at a fixed THz-field polarity. 
At operating points where such polarity reversals are possible, reliable polarity recovery requires phase stabilization at a known operating point or independent phase monitoring.

The difference in phase sensitivity can be seen directly from the
first-order response coefficient. Writing the differential signal as
\begin{equation}
    I_{\mathrm{diff}}
    =
    K_1(\varepsilon)\Delta\phi
    +\mathcal{O}\bigl((\Delta\phi)^3\bigr),
\end{equation}
we obtain the balanced-detection coefficient
\begin{equation}
    K_1^{\mathrm{bal}}
    =
    \frac{c\varepsilon_0E_0^2}{4}\cos 2\chi
    \left[
        \sqrt{1-4\delta^2}\cos\varepsilon-(1+2\delta)
    \right].
    \label{eq:K1bal}
\end{equation}
Therefore
\begin{equation}
    \left.
    \frac{\partial K_1^{\mathrm{bal}}}{\partial\varepsilon}
    \right|_{\varepsilon=0}
    =0.
\end{equation}

In contrast, for the phase-offset configuration,
\begin{equation}
    \begin{aligned}
    K_1^{\mathrm{PO}}
    &=
    -\frac{c\varepsilon_0E_0^2}{8}
    \left(1-10^{-\mathrm{OD}}\right)\sin 2\theta
    \\
    &\quad\times
    \Bigl[
        (1+2\delta)
        \\
    &\qquad\qquad
        -\sqrt{1-4\delta^2}
        \bigl(\cos\varepsilon-\cos 2\theta\sin\varepsilon\bigr)
    \Bigr],
    \end{aligned}
\end{equation}
which gives
\begin{equation}
    \left.
    \frac{\partial K_1^{\mathrm{PO}}}{\partial\varepsilon}
    \right|_{\varepsilon=0}
    =
    -\frac{c\varepsilon_0E_0^2}{16}
    \left(1-10^{-\mathrm{OD}}\right)
    \sqrt{1-4\delta^2}\,
    \sin 4\theta.
\end{equation}
Thus, for the small nonzero QWP rotation used here, the phase-offset
response varies linearly with small phase errors around
$\varepsilon=0$, whereas the balanced-detection response varies only
quadratically.
Consequently, small shot-to-shot phase fluctuations around $\varepsilon=0$ modulate the balanced-detection signal only in second order.
This first-order phase insensitivity is one reason for selecting
balanced detection for the near-dark-port operation considered below.

Balanced detection therefore combines a linear weak-field
response and cancellation of even-order waveform distortion
with a phase-stable first-order signal polarity on the
$\delta>0$ branch and first-order phase insensitivity of the
response coefficient at $\varepsilon=0$.

The signal-enhancement analysis presented below applies specifically to the QWP--PBS balanced-detection configuration in Fig.~\ref{Figure4}(b), rather than to the polarization-analysis-free configuration in Fig.~\ref{Figure2}(b).
In the balanced-detection configuration, a finite positive amplitude imbalance produces a first-order EO response even at zero phase imbalance, whereas the polarization-analysis-free configuration requires a finite phase bias for linear weak-field detection.
Thus, these two configurations demonstrate complementary capabilities of interferometric EO sampling: optical-phase readout without polarization analysis and enhanced, phase-robust linear detection under photodetector-saturation constraints.

\subsection{Signal enhancement factor}
\label{sec:enhancement}

We define the signal enhancement factor $G$ as the ratio of the
background-suppressed differential EO signal to the conventional
balanced-detection signal at the same nonzero EO retardation:
\begin{equation}
    G
    \equiv
    \frac{
    \left.I_{\mathrm{sig}}\right|_{\mathrm{BG\text{-}suppressed}}
    }{
    \left.I_{\mathrm{sig}}\right|_{\mathrm{conventional}}
    }.
    \label{eq:enhancement_definition}
\end{equation}
The conventional reference is obtained without a reference beam,
corresponding to $\delta=1/2$.
Its input power is adjusted until
$I_{\mathrm{tot}}=I_{\mathrm{sat}}$.
The resulting conventional signal is held fixed when evaluating
$G$ for different background-suppression settings and input powers.

Detector saturation is imposed as
\begin{equation}
I_{\mathrm{tot}}(\Delta\phi)
\le I_{\mathrm{sat}},
\label{eq:saturation-constraint}
\end{equation}
where $I_{\mathrm{sat}}$ corresponds to a saturation input power
$P_{\mathrm{sat}}=5~\mathrm{mW}$ per photodiode.
Because the probe and reference beams have identical spatial modes, all beams incident on the photodiodes share a common effective beam area $S$, and optical powers are related to intensities by $P=SI$; in particular, $P_{\mathrm{in}}=Sc\varepsilon_0E_0^2$, $P_{\mathrm{det}}(\Delta\phi)=SI_{\mathrm{det}}(\Delta\phi)$, and $P_{\mathrm{sat}}=SI_{\mathrm{sat}}$.
Because $G$ is a ratio of signals, $S$ cancels in the following results.

\begin{figure}[!t]
    \centering
    \includegraphics[width=\linewidth]{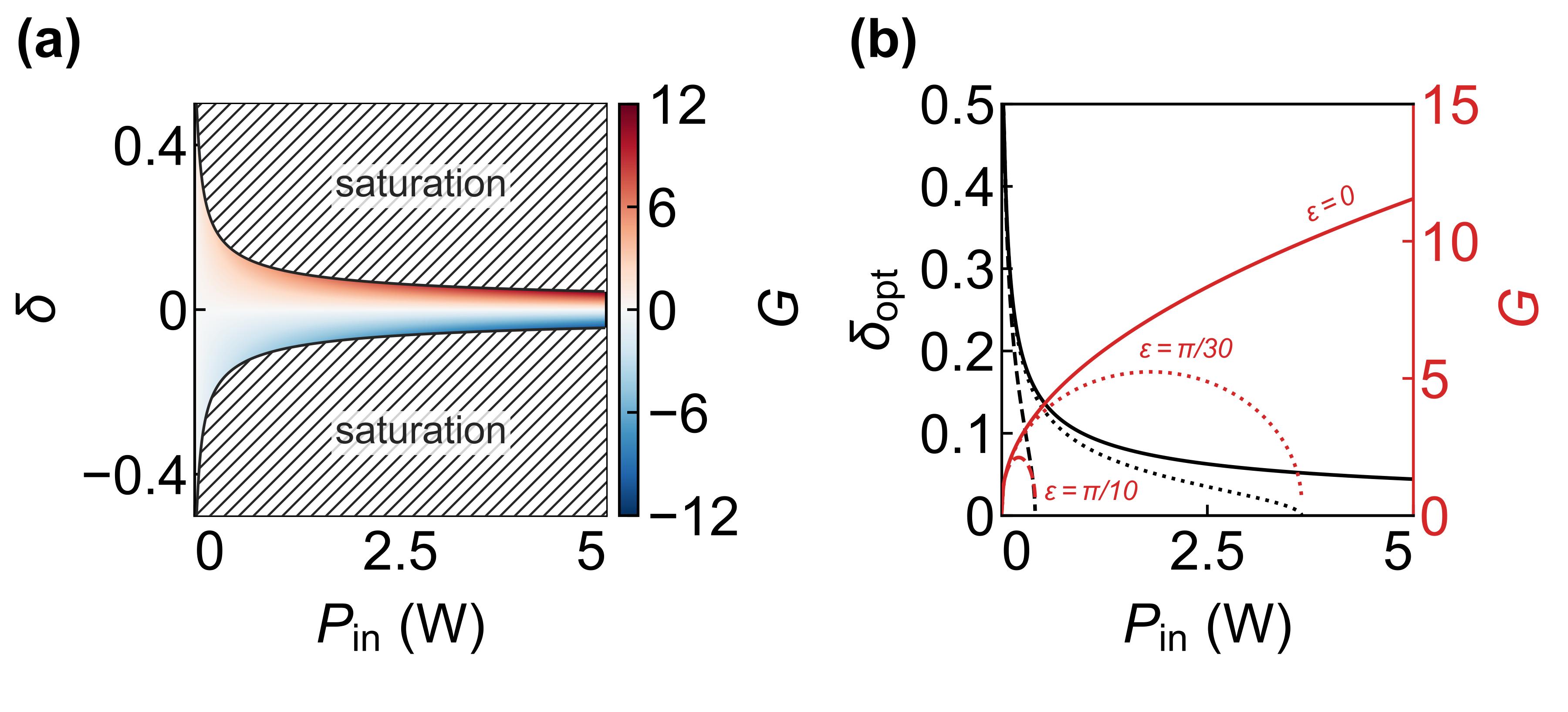}
    \caption{
    Signal enhancement in the balanced-detection configuration with a PBS extinction ratio of $\cot^2\chi=10^{3}$ under
    the detector-saturation constraint.
    (a) Enhancement factor $G$ as a function of the total input power $P_{\mathrm{in}}$ and
    amplitude imbalance $\delta$ for a phase imbalance of $\varepsilon=0$.
    The hatched regions indicate detector saturation for a saturation
    input power of $5~\mathrm{mW}$ on each photodiode.
    (b) Optimal amplitude imbalance $\delta_{\mathrm{opt}}$
    (black, left axis) and corresponding enhancement factor $G$
    (red, right axis) as functions of the total input power $P_{\mathrm{in}}$ for
    $\varepsilon=0$ (solid), $\varepsilon=\pi/30$ (dotted), and
    $\varepsilon=\pi/10$ (dashed).
    }
    \label{fig:enhancement}
\end{figure}

Figure~\ref{fig:enhancement}(a) shows the enhancement factor
as a function of $P_{\mathrm{in}}$ and $\delta$ at
$\varepsilon=0$; the hatched region denotes detector saturation.
For each $P_{\mathrm{in}}$, we optimize $G$ over the
$0<\delta\leq1/2$ branch subject to
Eq.~(\ref{eq:saturation-constraint}).
The behavior of the optimum can be obtained analytically in the
weak-field limit.
At $\varepsilon=0$, the first-order enhancement factor is
\begin{equation}
G(\delta)
=
\frac{P_{\mathrm{in}}}{8P_{\mathrm{sat}}}
\left[
1+2\delta-\sqrt{1-4\delta^2}
\right],
\label{eq:G_weak}
\end{equation}
where the conventional balanced-detection reference is operated
at the same per-photodiode saturation limit.
Since $G$ increases monotonically for $0<\delta<1/2$, the optimum is the largest $\delta$ allowed by detector saturation.
Equation~(\ref{eq:G_weak}) and the weak-field results below do not depend on $\chi$, because the factor $\cos 2\chi$ is common to the background-suppressed and conventional signals, which use the same PBS, and the zero-field photodiode intensities are independent of $\chi$.

The zero-field background power on each photodiode is
\begin{equation}
P_{\mathrm{det}}(\Delta\phi=0)
=
\frac{P_{\mathrm{in}}}{4}
\left[
1-\sqrt{1-4\delta^2}
\right].
\label{eq:Pdet_background}
\end{equation}
Thus, in the saturation-limited regime,
$P_{\mathrm{det}}(\Delta\phi=0)=P_{\mathrm{sat}}$ gives
\begin{equation}
\delta_{\mathrm{opt}}
=
\frac{1}{2}
\sqrt{
1-
\left(
1-\frac{4P_{\mathrm{sat}}}{P_{\mathrm{in}}}
\right)^2
},
\label{eq:delta_opt}
\end{equation}
and hence
\begin{equation}
G_{\mathrm{opt}}
=
\frac{1}{2}
+
\sqrt{
\frac{P_{\mathrm{in}}}{8P_{\mathrm{sat}}}
-\frac{1}{4}
},
\label{eq:G_opt}
\end{equation}
which gives the $\sqrt{P_{\mathrm{in}}}$ scaling observed in
$G$.

Because both photodiodes operate near $P_\mathrm{sat}$ at the optimum, 
the detected photocurrents and hence the shot noise of the
differential output are comparable in the two cases.
In the shot-noise-limited regime, the signal enhancement $G$ therefore translates
into a corresponding improvement in the signal-to-noise ratio.

Figure~\ref{fig:enhancement}(b) shows $\delta_{\mathrm{opt}}$
and the corresponding enhancement factor for
$\varepsilon=0$, $\pi/30$, and $\pi/10$.
At $\varepsilon=0$, the optimized enhancement scales
approximately with $\sqrt{P_{\mathrm{in}}}$ in the weak-field,
strong-background-suppression regime where the residual
background remains much larger than the EO-induced contribution
at the detector.
At $P_{\mathrm{in}}=5~\mathrm{W}$, the calculated differential
signal is approximately 12 times the conventional signal at
the same per-photodiode saturation limit.

For finite phase imbalance, a residual background remains even at
$\delta=0$ and increases proportionally with $P_{\mathrm{in}}$.
As $P_{\mathrm{in}}$ increases, this irreducible background consumes
an increasing fraction of the available detector dynamic range.
Consequently, the maximum allowable positive amplitude imbalance
must be reduced to avoid saturation, which in turn reduces the
linear EO signal.
At sufficiently high input power, the residual background alone
exceeds the detector saturation limit, so that no value of $\delta$
on the selected branch is admissible.

\subsection{Waveform distortion and linear-response range}

Interferometric background suppression enhances the differential
EO signal under detector-saturation constraints, but the residual
nonlinear response limits the usable field range.
As shown in Eq.~(\ref{eq:balanced_difference}), all even-order
contributions cancel in the balanced-detection output, so the
leading nonlinear correction is cubic at operating points with
a nonzero first-order coefficient.
Here, we evaluate this readout nonlinearity at zero phase
imbalance, $\varepsilon=0$, on the $0<\delta\leq1/2$ branch.

At fixed input power and nonzero EO retardation, the magnitude
of $I^{(1)}$ approaches zero as $\delta$ approaches zero,
whereas the cubic-response coefficient remains finite.
The relative cubic contribution therefore increases as the
background-cancellation point is approached, narrowing the
range of retardations over which the response remains
approximately linear.

We adopt a $1\%$ criterion for the relative cubic contribution:
\begin{equation}
    \left|
    \frac{I^{(3)}}{I^{(1)}}
    \right|
    \leq 0.01.
    \label{eq:linear_response_criterion}
\end{equation}
For a sampled waveform, this criterion is evaluated at the
maximum absolute retardation, with the interferometer settings
held fixed.
From Eq.~(\ref{eq:balanced_difference}), the ratio at
$\varepsilon=0$ is
\begin{equation}
    \left|
    \frac{I^{(3)}}{I^{(1)}}
    \right|
    =
    \left|
    \frac{
    4(1+2\delta)-\sqrt{1-4\delta^2}
    }{
    24\left[
    \sqrt{1-4\delta^2}-(1+2\delta)
    \right]
    }
    \right|
    (\Delta\phi)^2.
    \label{eq:third_to_first_ratio}
\end{equation}

At fixed $\delta$ and $\Delta\phi$, this ratio is independent
of the total input power $P_{\mathrm{in}}$ within the present
model, because the first- and third-order contributions have
the same overall dependence on optical power.
However, the relative nonlinear contribution can vary with
input power when $\delta$ is adjusted to its power-dependent
optimum under the detector-saturation constraint.

For the retardation magnitude used in the numerical calculations,
$|\Delta\phi|=2.2\times10^{-3}~\mathrm{rad}$, the $1\%$
criterion is satisfied over the range
\begin{equation}
    3.0\times10^{-5}\lesssim\delta\leq1/2.
    \label{eq:linear_response_delta_range}
\end{equation}

In particular, the optimized zero-phase-imbalance operating
point in Fig.~\ref{fig:enhancement}(b) at
$P_{\mathrm{in}}=5~\mathrm{W}$ and
$P_{\mathrm{sat}}=5~\mathrm{mW}$ per photodiode is
$\delta_{\mathrm{opt}}\simeq4.4\times10^{-2}$.
Evaluating Eq.~(\ref{eq:third_to_first_ratio}) at this operating
point gives
$|I^{(3)}/I^{(1)}|\simeq7.2\times10^{-6}$.
Thus, the approximately 12-fold signal enhancement shown in
Fig.~\ref{fig:enhancement}(b) is obtained with a cubic
contribution well below the $1\%$ criterion, demonstrating
that substantial signal enhancement and an approximately
linear EO response can be achieved simultaneously within
the present model.

\section{Conclusion}
In this work, we developed a Jones-matrix framework for background-suppressed EO sampling based on destructive interference between an EO-modulated probe and an unmodulated reference beam.
Among the five configurations considered, balanced detection provides a linear weak-field response at finite positive amplitude imbalance, whereas exact background cancellation eliminates the linear term and leaves a cubic leading response.
Its differential output cancels all even-order contributions, avoiding second-order waveform distortion.
Moreover, on the positive-amplitude-imbalance branch, the first-order signal retains its polarity against phase-imbalance variations and is insensitive to small phase errors to first order around zero phase imbalance.

Under a per-photodiode saturation constraint, the signal is maximized by adjusting the amplitude imbalance so that the detector input approaches the saturation limit without exceeding it.
At zero phase imbalance, the optimized enhancement scales approximately with the square root of the total probe-plus-reference input power.
For a total input power of $5~\mathrm{W}$ and a saturation limit of $5~\mathrm{mW}$ per photodiode, the calculated differential signal is approximately 12 times that of conventional balanced detection.
At this operating point, the cubic contribution is only approximately $7.2\times10^{-6}$ times that of the linear contribution for $|\Delta\phi|=2.2\times10^{-3}~\mathrm{rad}$.
Finite phase imbalance leaves residual background, which consumes part of the available detector dynamic range and thereby reduces the achievable enhancement.

Interferometric suppression can also act multiplicatively with polarization filtering, further reducing background transmitted through a finite-extinction polarizer.
The framework may also be applicable to other background-suppressed polarimetric measurements of weak phase retardations, including vacuum birefringence, axion-induced polarization effects, and weak magneto-optical signals.

Beyond sensitivity enhancement, interferometric EO detection provides access to optical phase shifts relative to an independent reference beam, even when the probe polarization state remains unchanged, as demonstrated by the configuration in Fig.~\ref{Figure2}(b).
This capability could, in principle, be extended to EO-induced phase shifts produced by the Pockels effect in LiNbO$_3$ under a configuration that preserves the probe polarization state~\cite{Weis1985,Wang2018LN}.
LiNbO$_3$ is widely used for electro-optic modulators and integrated photonics, and such phase shifts could enable electric-field detection even in configurations where conventional polarization-based readout provides no signal.
Furthermore, combining common-mode and differential phase measurements with appropriate detection channels could enable simultaneous reconstruction of two orthogonal electric-field components, with potential applications in vector-resolved THz field measurements and THz ellipsometry~\cite{Lenz2025}.
These extensions are beyond the scope of the present model and should be distinguished from the sensitivity enhancement demonstrated for the balanced-detection configuration in Fig.~\ref{Figure4}(b).

\begin{backmatter}
\bmsection{Funding} 
This work was supported by JSPS KAKENHI Grant Numbers JP25K17370, JP25K17965 and JP25K21664, and Japan Science and Technology Agency PRESTO JPMJPR24L5.

\bmsection{Disclosures} 
The authors declare no conflicts of interest.

\bmsection{Data Availability Statement} 
Data underlying the results presented in this paper are not publicly available at this time but may be obtained from the authors upon reasonable request.

\end{backmatter}
\bibliography{Reference}

@article{Valdmanis1983,
  article_type = {journal},
  title = {Subpicosecond electrical sampling},
  author = {Valdmanis, J. A. and Mourou, G. A. and Gabel, C. W.},
  journal = {IEEE Journal of Quantum Electronics},
  volume = {19},
  number = {4},
  pages = {664--667},
  year = {1983},
  doi = {10.1109/JQE.1983.1071915}
}

@article{Auston1984,
  article_type = {journal},
  title = {Cherenkov Radiation from Femtosecond Optical Pulses in Electro-Optic Media},
  author = {Auston, D. H. and Cheung, K. P. and Valdmanis, J. A. and Kleinman, D. A.},
  journal = {Physical Review Letters},
  volume = {53},
  number = {16},
  pages = {1555--1558},
  year = {1984},
  doi = {10.1103/PhysRevLett.53.1555}
}

@article{Wu1995,
  article_type = {journal},
  title = {Free-space electro-optic sampling of terahertz beams},
  author = {Wu, Q. and Zhang, X.-C.},
  journal = {Applied Physics Letters},
  volume = {67},
  number = {24},
  pages = {3523--3525},
  year = {1995},
  doi = {10.1063/1.114909}
}

@article{Duvillaret1996,
  article_type = {journal},
  title = {A reliable method for extraction of material parameters in terahertz time-domain spectroscopy},
  author = {Duvillaret, L. and Garet, F. and Coutaz, J.-L.},
  journal = {IEEE Journal of Selected Topics in Quantum Electronics},
  volume = {2},
  number = {3},
  pages = {739--746},
  year = {1996},
  doi = {10.1109/2944.571775}
}

@article{Koch2023,
  article_type = {journal},
  title = {Terahertz time-domain spectroscopy},
  author = {Koch, Martin and Mittleman, Daniel M. and Ornik, Jan and Castro-Camus, Enrique},
  journal = {Nature Reviews Methods Primers},
  volume = {3},
  number = {1},
  pages = {48},
  year = {2023},
  doi = {10.1038/s43586-023-00232-z}
}

@article{Pupeza2020,
  article_type = {journal},
  title = {Field-resolved infrared spectroscopy of biological systems},
  author = {Pupeza, Ioachim and Huber, Marinus and Trubetskov, Michael and Schweinberger, Wolfgang and Hussain, Syed A. and Hofer, Christina and Fritsch, Kilian and Poetzlberger, Markus and Vamos, Lenard and Fill, Ernst and Amotchkina, Tatiana and Kepesidis, Kosmas V. and Apolonski, Alexander and Karpowicz, Nicholas and Pervak, Vladimir and Pronin, Oleg and Fleischmann, Frank and Azzeer, Abdallah and {\v{Z}}igman, Mihaela and Krausz, Ferenc},
  journal = {Nature},
  volume = {577},
  number = {7788},
  pages = {52--59},
  year = {2020},
  doi = {10.1038/s41586-019-1850-7}
}

@article{Okazaki2026,
author = {Daiki Okazaki and Max Koch and Mojtaba Aghakasiri and Philipp Steinleitner and Wolfgang Schweinberger and Dionysios Potamianos and Yang Gui and Aleksandar Sebesta and Behnam Jahedi Abbasvand and Vipulkumar Bhaliya and Shigeki Tokita and Ferenc Krausz and Alexander Weigel},
journal = {Optics Express},
number = {16},
pages = {28928--28942},
title = {1-{THz}-to-mid-infrared field-resolved spectrometer driven by a {Cr:ZnS} oscillator},
volume = {34},
month = {Aug},
year = {2026},
doi = {10.1364/OE.596193}
}

@article{Kampfrath2013,
  article_type = {journal},
  title = {Terahertz spin current pulses controlled by magnetic heterostructures},
  author = {Kampfrath, T. and Battiato, M. and Maldonado, P. and Eilers, G. and N{\"o}tzold, J. and M{\"a}hrlein, S. and Zbarsky, V. and Freimuth, F. and Mokrousov, Y. and Bl{\"u}gel, S. and Wolf, M. and Radu, I. and Oppeneer, P. M. and M{\"u}nzenberg, M.},
  journal = {Nature Nanotechnology},
  volume = {8},
  number = {4},
  pages = {256--260},
  year = {2013},
  doi = {10.1038/nnano.2013.43}
}

@article{Sotome2021,
  article_type = {journal},
  title = {Terahertz emission spectroscopy of ultrafast exciton shift current in the noncentrosymmetric semiconductor {CdS}},
  author = {Sotome, M. and Nakamura, M. and Morimoto, T. and Zhang, Y. and Guo, G.-Y. and Kawasaki, M. and Nagaosa, N. and Tokura, Y. and Ogawa, N.},
  journal = {Physical Review B},
  volume = {103},
  number = {24},
  pages = {L241111},
  year = {2021},
  doi = {10.1103/PhysRevB.103.L241111}
}

@article{Pompili2018,
  article_type = {journal},
  title = {Ultrafast evolution of electric fields from high-intensity laser--matter interactions},
  author = {Pompili, R. and Anania, M. P. and Bisesto, F. and Botton, M. and Chiadroni, E. and Cianchi, A. and Curcio, A. and Ferrario, M. and Galletti, M. and Henis, Z. and Petrarca, M. and Schleifer, E. and Zigler, A.},
  journal = {Scientific Reports},
  volume = {8},
  number = {1},
  pages = {3243},
  year = {2018},
  doi = {10.1038/s41598-018-21711-4}
}

@article{Riek2015,
  article_type = {journal},
  title = {Direct sampling of electric-field vacuum fluctuations},
  author = {Riek, C. and Seletskiy, D. V. and Moskalenko, A. S. and Schmidt, J. F. and Krauspe, P. and Eckart, S. and Eggert, S. and Burkard, G. and Leitenstorfer, A.},
  journal = {Science},
  volume = {350},
  number = {6259},
  pages = {420--423},
  year = {2015},
  doi = {10.1126/science.aac9788}
}

@article{Berden2007,
  article_type = {journal},
  title = {Benchmarking of Electro-Optic Monitors for Femtosecond Electron Bunches},
  author = {Berden, G. and Gillespie, W. A. and Jamison, S. P. and Knabbe, E.-A. and MacLeod, A. M. and van der Meer, A. F. G. and Phillips, P. J. and Schlarb, H. and Schmidt, B. and Schm{\"u}ser, P. and Steffen, B.},
  journal = {Physical Review Letters},
  volume = {99},
  number = {16},
  pages = {164801},
  year = {2007},
  doi = {10.1103/PhysRevLett.99.164801}
}

@article{Ota2021,
  article_type = {journal},
  title = {Longitudinal and transverse spatial beam profile measurement of relativistic electron bunch by electro-optic sampling},
  author = {Ota, Masato and Kan, Koichi and Komada, Soichiro and Arikawa, Yasunobu and Shimizu, Tomoki and Mag-usara, Valynn Katrine and Sakawa, Youichi and Matsui, Tatsunosuke and Nakajima, Makoto},
  journal = {Applied Physics Express},
  volume = {14},
  number = {2},
  pages = {026503},
  year = {2021},
  doi = {10.35848/1882-0786/abd867}
}

@article{Ota2022,
  article_type = {journal},
  title = {Ultrafast visualization of an electric field under the {Lorentz} transformation},
  author = {Ota, Masato and Kan, Koichi and Komada, Soichiro and Wang, Youwei and Agulto, Verdad C. and Mag-usara, Valynn Katrine and Arikawa, Yasunobu and Asakawa, Makoto R. and Sakawa, Youichi and Matsui, Tatsunosuke and Nakajima, Makoto},
  journal = {Nature Physics},
  volume = {18},
  number = {12},
  pages = {1436--1440},
  year = {2022},
  doi = {10.1038/s41567-022-01767-w}
}

@article{BeneaChelmus2025,
  article_type = {journal},
  title = {Electro-optic sampling of classical and quantum light},
  author = {Benea-Chelmus, Ileana-Cristina and Faist, J{\'e}r{\^o}me and Leitenstorfer, Alfred and Moskalenko, Andrey S. and Pupeza, Ioachim and Seletskiy, Denis V. and Vodopyanov, Konstantin L.},
  journal = {Optica},
  volume = {12},
  number = {4},
  pages = {546--563},
  year = {2025},
  doi = {10.1364/OPTICA.544333}
}

@article{Huang2024,
  article_type = {journal},
  title = {Electro-optic {3D} snapshot of a laser wakefield accelerated kilo-ampere electron bunch},
  author = {Huang, Kai and Jin, Zhan and Nakanii, Nobuhiko and Hosokai, Tomonao and Kando, Masaki},
  journal = {Light: Science \& Applications},
  volume = {13},
  number = {1},
  pages = {84},
  year = {2024},
  doi = {10.1038/s41377-024-01440-2}
}

@article{Pan2014,
  article_type = {journal},
  title = {Modified {THz} electro-optic sampling for high optical modulation depth, large dynamical range, and low background noises},
  author = {Pan, Xinjian and Cai, Yi and Zeng, Xuanke and Lu, Xiaowei and Zhang, Dongping and Li, Jingzhen and Chen, Hongyi and Xu, Shixiang},
  journal = {Optics Letters},
  volume = {39},
  number = {13},
  pages = {3778--3781},
  year = {2014},
  doi = {10.1364/OL.39.003778}
}

@article{Porer2014,
  article_type = {journal},
  title = {Shot noise reduced terahertz detection via spectrally postfiltered electro-optic sampling},
  author = {Porer, Michael and M{\'e}nard, Jean-Michel and Huber, Rupert},
  journal = {Optics Letters},
  volume = {39},
  number = {8},
  pages = {2435--2438},
  year = {2014},
  doi = {10.1364/OL.39.002435}
}

@article{Jiang1999,
  article_type = {journal},
  title = {Electro-optic sampling near zero optical transmission point},
  author = {Jiang, Zhiping and Sun, F. G. and Chen, Q. and Zhang, X.-C.},
  journal = {Applied Physics Letters},
  volume = {74},
  number = {9},
  pages = {1191--1193},
  year = {1999},
  doi = {10.1063/1.123495}
}

@article{Xu2008,
  article_type = {journal},
  title = {A theoretical and experimental research on terahertz electro-optic sampling at near-zero optical transmission point},
  author = {Xu, Shi-Xiang and Cai, Hua},
  journal = {Chinese Physics Letters},
  volume = {25},
  number = {1},
  pages = {152--155},
  year = {2008},
  doi = {10.1088/0256-307X/25/1/042}
}

@article{Brunner2014,
  article_type = {journal},
  title = {Distortion-free enhancement of terahertz signals measured by electro-optic sampling. {I}. {Theory}},
  author = {Brunner, Fabian D. J. and Johnson, Jeremy A. and Gr{\"u}bel, Sebastian and Ferrer, Andr{\'e}s and Johnson, Steven L. and Feurer, Thomas},
  journal = {Journal of the Optical Society of America B},
  volume = {31},
  number = {4},
  pages = {904--910},
  year = {2014},
  doi = {10.1364/JOSAB.31.000904}
}

@article{Johnson2014,
  article_type = {journal},
  title = {Distortion-free enhancement of terahertz signals measured by electro-optic sampling. {II}. {Experiment}},
  author = {Johnson, Jeremy A. and Brunner, Fabian D. J. and Gr{\"u}bel, Sebastian and Ferrer, Andr{\'e}s and Johnson, Steven L. and Feurer, Thomas},
  journal = {Journal of the Optical Society of America B},
  volume = {31},
  number = {5},
  pages = {1035--1040},
  year = {2014},
  doi = {10.1364/JOSAB.31.001035}
}

@inproceedings{Kato2018Polarization,
  article_type = {conference},
  title = {Enhancement of {THz} {EO} Sampling Signal by Polarization Filtering},
  author = {Kato, Hiroyuki and Kitahara, Hideaki and Goto, Daiki and Yasumoto, Takuro and Shihara, Masaki and Afalla, Jessica and Mag-usara, Valynn and Bulgarevich, Dmitry and Esca{\~n}o, Clare and Yamamoto, Kohji and Furuya, Takashi and Bakunov, Michael and Estacio, Elmer and Tani, Masahiko},
  booktitle = {2018 43rd International Conference on Infrared, Millimeter, and Terahertz Waves ({IRMMW-THz})},
  year = {2018},
  publisher = {IEEE},
  doi = {10.1109/IRMMW-THz.2018.8510458}
}

@article{Stromberg2024,
  author  = {Christopher J. Stromberg and Edwin J. Heilweil},
  title   = {Practical considerations for the amplification of electro-optically detected {THz} signals},
  journal = {J. Opt. Soc. Am. B},
  volume  = {41},
  number  = {10},
  pages   = {2350--2357},
  year    = {2024},
  doi     = {10.1364/JOSAB.535012}
}

@article{Beckh2021,
  article_type = {journal},
  title = {Analysis of Subcycle Electro-Optic Sampling Without Background},
  author = {Beckh, Cornelius and Sulzer, Philipp and Fritzsche, Niklas and Riek, Claudius and Leitenstorfer, Alfred},
  journal = {Journal of Infrared, Millimeter, and Terahertz Waves},
  volume = {42},
  number = {6},
  pages = {701--714},
  year = {2021},
  doi = {10.1007/s10762-021-00789-4}
}

@article{Abbott2016,
  article_type = {journal},
  title = {Observation of Gravitational Waves from a Binary Black Hole Merger},
  author = {Abbott, B. P. and others},
  journal = {Physical Review Letters},
  volume = {116},
  number = {6},
  pages = {061102},
  year = {2016},
  doi = {10.1103/PhysRevLett.116.061102}
}

@article{Tomberg2019,
  article_type = {journal},
  title = {Background-free broadband absorption spectroscopy based on interferometric suppression with a sign-inverted waveform},
  author = {Tomberg, Teemu and Muraviev, Andrey and Ru, Qitian and Vodopyanov, Konstantin L.},
  journal = {Optica},
  volume = {6},
  number = {2},
  pages = {147--151},
  year = {2019},
  doi = {10.1364/OPTICA.6.000147}
}

@article{Song2022,
  article_type = {journal},
  title = {Broadband background-free vibrational spectroscopy using a mode-locked {Cr:ZnS} laser},
  author = {Song, Wenqing and Okazaki, Daiki and Morichika, Ikki and Ashihara, Satoshi},
  journal = {Optics Express},
  volume = {30},
  number = {21},
  pages = {38674--38683},
  year = {2022},
  doi = {10.1364/OE.470893}
}

@article{Clough2009,
  article_type = {journal},
  title = {Detection of terahertz pulses using a modified {Sagnac} interferometer},
  author = {Clough, Benjamin and Hurley, David H. and Han, Pengyu and Liao, Jun and Huang, Rena and Zhang, X.-C.},
  journal = {Sensing and Imaging: An International Journal},
  volume = {10},
  number = {3--4},
  pages = {55--62},
  year = {2009},
  doi = {10.1007/s11220-009-0046-x}
}

@article{Ibrahim2016,
  article_type = {journal},
  title = {Ultra-high dynamic range electro-optic sampling for detecting millimeter and sub-millimeter radiation},
  author = {Ibrahim, Akram and F{\'e}rachou, Denis and Sharma, Gargi and Singh, Kanwarpal and Kirouac-Turmel, Marie and Ozaki, Tsuneyuki},
  journal = {Scientific Reports},
  volume = {6},
  number = {1},
  pages = {23107},
  year = {2016},
  doi = {10.1038/srep23107}
}

@article{Wilke2024,
  article_type = {journal},
  title = {Thin-film lithium niobate electro-optic terahertz wave detector},
  author = {Wilke, Ingrid and Monahan, Jackson and Toroghi, Seyfollah and Rabiei, Payam and Hine, George},
  journal = {Scientific Reports},
  volume = {14},
  number = {1},
  pages = {4822},
  year = {2024},
  doi = {10.1038/s41598-024-55156-9}
}

@article{Gallot1999,
  article_type = {journal},
  title = {Electro-optic detection of terahertz radiation},
  author = {Gallot, G. and Grischkowsky, D.},
  journal = {Journal of the Optical Society of America B},
  volume = {16},
  number = {8},
  pages = {1204--1212},
  year = {1999},
  doi = {10.1364/JOSAB.16.001204}
}

@article{Casalbuoni2008,
  article_type = {journal},
  title = {Numerical studies on the electro-optic detection of femtosecond electron bunches},
  author = {Casalbuoni, S. and Schlarb, H. and Schmidt, B. and Schm{\"u}ser, P. and Steffen, B. and Winter, A.},
  journal = {Physical Review Special Topics--Accelerators and Beams},
  volume = {11},
  number = {7},
  pages = {072802},
  year = {2008},
  doi = {10.1103/PhysRevSTAB.11.072802}
}

@article{Marple1964,
  author    = {Marple, D. T. F.},
  title     = {Refractive Index of {ZnSe}, {ZnTe}, and {CdTe}},
  journal   = {J. Appl. Phys.},
  volume    = {35},
  number    = {3},
  pages     = {539--542},
  month     = {Mar},
  year      = {1964},
  publisher = {AIP Publishing},
  doi       = {10.1063/1.1713411},
  url       = {https://doi.org/10.1063/1.1713411}
}

@article{Weis1985,
  author  = {R. S. Weis and T. K. Gaylord},
  title   = {Lithium niobate: Summary of physical properties and crystal structure},
  journal = {Applied Physics A},
  volume  = {37},
  number  = {4},
  pages   = {191--203},
  year    = {1985},
  doi     = {10.1007/BF00614817}
}

@article{Wang2018LN,
  author  = {Cheng Wang and Mian Zhang and Xi Chen and Maxime Bertrand and Amirhassan Shams-Ansari and Sethumadhavan Chandrasekhar and Peter Winzer and Marko Loncar},
  title   = {Integrated lithium niobate electro-optic modulators operating at {CMOS}-compatible voltages},
  journal = {Nature},
  volume  = {562},
  number  = {7725},
  pages   = {101--104},
  year    = {2018},
  doi     = {10.1038/s41586-018-0551-y}
}

@article{Lenz2025,
  author  = {Maximilian Lenz and Pietro Musumeci},
  title   = {Single-shot electro-optic sampling with arbitrary terahertz polarization},
  journal = {Photonics Research},
  volume  = {13},
  number  = {6},
  pages   = {1736--1746},
  year    = {2025},
  doi     = {10.1364/PRJ.558770}
}
\end{document}